\documentclass[prd,showpacs,preprintnumbers,amsmath,amssymb]{revtex4}
\usepackage{graphicx}% Include figure files
\usepackage{dcolumn}% Align table columns on decimal point
\usepackage{bm}% bold math

\begin{document}

\title{Equivalence between the phase-integral and worldline-instanton methods}
% for Schwinger pair production by electric fields that depend on one time or space coordinate}

\author{Sang Pyo Kim}\email{E-mail: sangkim@kunsan.ac.kr}
\affiliation{Department of Physics, Kunsan National University, Kunsan 54150, Korea\footnote{Permanent address}}
\affiliation{Center for Relativistic Laser Science, Institute for Basic Science, Gwangju 61005, Korea}
\affiliation{Institute of Theoretical Physics, Chinese Academy of Sciences, Beijing 100190, China}
\author{Don N. Page} \email{E-mail:profdonpage@gmail.com}
\affiliation{Department of Physics,Theoretical Physics Institute, 4-181 CCIS, University of Alberta, Edmonton, Alberta T6G 2E1, Canada}

\date{\today}

\begin{abstract}
The phase-integral and worldline-instanton methods are two widely used methods to calculate Schwinger pair-production densities in electric fields of fixed direction that depend on just one time or space coordinate in the same fixed plane of the electromagnetic field tensor.  We show that for charged spinless bosons the leading results of the phase-integral method integrated up through quadratic momenta are equivalent to those of the worldline-instanton method including prefactors. We further apply the phase-integral method to fermion production and time-dependent electric fields parallel to a constant magnetic field.
\end{abstract}


\maketitle

\section{INTRODUCTION}

Quantum field theory predicts that a strong electromagnetic field not only polarizes the vacuum but also creates electron-positron pairs as a consequence of interactions of photons and virtual electrons in the Dirac sea. The so-called Schwinger pair production first discovered by Sauter and comprehensively formulated by Schwinger is the result of nonperturbative quantum effects \cite{Sauter:1931zz,Schwinger:1951nm}. The rapid development of high intense lasers via chirped pulse amplification technology may open a window for direct observations of electron-positron pair production in the near future (for review and references, see Refs.~\cite{Dunne:2008kc,DiPiazza:2011tq}).
%On the other hand, there are many astrophysical sources for strong electromagnetic fields, such as neutron stars or magnetars, and the Schwinger pair production has been suggested as a mechanism for producing gamma rays in astrophysics (for review and references, see Ref.~\cite{Ruffini:2009hg}).

The Dirac equation or Klein-Gordon equation for charges interacting with external electromagnetic fields, however, has
not yet been solved exactly except for some simple profiles of electric or magnetic fields, which do not include pulsed focused electromagnetic fields from colliding laser beams.
Hence intensive investigations have been focussed on one-dimensional electric fields to see the pulse effect of temporal or spatial localizations. A few field theoretical methods have been proposed to obtain analytical expressions for the leading contributions to Schwinger pair production in one-dimensional time-dependent or space-dependent electric fields. One of them is the worldline instanton method, in which a scalar charge in an electric field undergoes periodic motion in the Euclidean spacetime or space and the relativistic action of the periodic orbit gives the worldline instanton for pair production \cite{Dunne:2005sx,Dunne:2006st}. The phase-integral method looks for imaginary contributions in the complex plane of time or space \cite{Kim:2000un,Kim:2003qp,Kim:2007pm} to the relativistic action from the field equations and the leading Wentzel-Kramers-Brillouin (WKB) action in electric fields. One of the interesting results of the phase-integral method is the Stokes phenomenon in some electric field profiles, by which the mean number of produced pairs exhibits substructure \cite{Dumlu:2010ua,Dumlu:2011rr,Dumlu:2011cc}. Some other WKB methods were employed to explain the Schwinger pair production \cite{Brezin:1970xf,Marinov:1977gq,Gordon:2014aba,Linder:2015vta,Torgrimsson:2016ant,Rajeev:2017smn,Linder:2018mwk} and the residue theorem was applied to calculate the mean number from the frequency in complex time \cite{Kim:2013iua,Kim:2013cka}.

In this paper we show that in an electric field in a fixed direction and with a value depending on one time or space coordinate in the same timelike 2-plane of electromagnetic field tensor, the leading terms of Schwinger pair production densities from the phase-integral method \cite{Kim:2007pm} give those from the worldline instantons and their prefactors \cite{Dunne:2005sx,Dunne:2006st}.
The reason for this equivalence is that the WKB action from the field equation carries the quantum fluctuations in the transverse plane as well as in the longitudinal direction, while the worldline instanton is the action along the field in the Euclidean spacetime and the prefactor takes into account fluctuations. We show that for any one-dimensional electric field, either time-dependent or space-dependent in the restricted sense given above, the mean number from the WKB action of the phase-integral when integrated over the quadratic momenta recovers the pair production density from the worldline instanton with the prefactor included. The momenta integrals beyond quadratic order, when expanded in a power series, provide corrections to the worldline instanton results. We further show that the phase-integral method for fermion production in a Sauter-type electric field \cite{Kim:2007pm} can be extended to any generic one-dimensional electric field and to boson production in time-dependent electric fields parallel to a constant magnetic field.

The organization of this paper is as follows. In Sec.~\ref{sec2}, for one-dimensional, time-dependent or space-dependent electric fields, we show the equivalence of the pair production densities of scalars between the phase-integral method and the worldline instanton method. In Sec.~\ref{sec3}, we extend the phase-integral method to the pair production densities of fermions in time-dependent electric fields. In Sec.~\ref{sec4}, we find the pair production densities in time-dependent electric fields combined with a parallel constant magnetic field. In %Sec.~\ref{sec5}, we discuss the physics related to the phase-integral in the case of localized pulsed electromagnetic fields from collision of intense lasers.

\section{Equivalence of scalar pair production}\label{sec2}

A charged scalar field with mass $m$ and charge $q$ in a spatially homogeneous but time-dependent electric field with the vector potential
\begin{eqnarray}
E_{\parallel} (t) = - \dot{A}_{\parallel} (t)
\end{eqnarray}
obeys the Fourier-momentum mode equation (in natural units with $c = \hbar = 4 \pi \epsilon_0 = 1$)
\begin{eqnarray}
\ddot{\phi}_{\bf k} (t) + \omega_{\bf k}^2 (t) \phi_{\bf k} (t) = 0, \label{sc eq}
\end{eqnarray}
where
\begin{eqnarray}
\omega_{\bf k}^2 (t) =  m^2 + {\bf k}_{\perp}^2 + \bigl(k_{\parallel} - qA_{\parallel} (t) \bigr)^2. \label{t-freq}
\end{eqnarray}
In natural units $E$ and $m^2$ have the same dimension of inverse squared length or time. Note that Eq. (\ref{sc eq}) describes a problem of scattering in a potential barrier. For generic electric fields a positive-frequency solution of Eq.~(\ref{sc eq}) in the remote past splits into another positive-frequency solution and a negative-frequency solution in the remote future, and the relative ratio of the coefficient of the negative-frequency solution in the remote future to that of the positive-frequency solution in the remote past gives the Bogoliubov coefficient for pair production.

In the phase-integral method, the phase $S_{\bf k}(t)$ of the field $\phi_{\bf k} = e^{- i S_{\bf k}(t)}$ can obtain an imaginary part due to spontaneous pair production. The density of pair production is the momentum sum of pairs with ${\bf k}$, and another sum of $-{\bf k}$ gives the same density of pair production. Thus, taking the square root of the product of the sum of ${\bf k}$ and another sum of $-{\bf k}$ and using the WKB action to the phase, the density of produced pairs of Ref.~\cite{Kim:2007pm} is approximately given by
\begin{eqnarray}
\frac{d{\cal N}_{\bf k}}{dx^3} = \int \frac{d^{3} {\bf k}}{(2 \pi)^3} e^{- i \oint \bar{\omega}_{\bf k} (z) dz},
\end{eqnarray}
where the time is continued to a complex variable $z$ and $\bar{\omega}_{\bf k}$ is the average of the frequencies with ${\bf k}$ and $-{\bf k}$ in Eq.~(\ref{t-freq}),
\begin{eqnarray}
\bar{\omega}_{\bf k} (z) = \frac{1}{2} \Bigl(\omega_{\bf k} (z) + \omega_{-{\bf k}} (z) \Bigr).
\end{eqnarray}
Here, $\oint$ means either a contour integral in the complex time plane $z$ after taking appropriate branch cuts \cite{Kim:2007pm} or a complete integral for the back and forth motion in the Euclidean time \cite{Dunne:2005sx,Dunne:2006st}.

We assume that the variation of the vector potential is characterized by a parameter $T$, the inverse frequency in an oscillating field, such that
\begin{eqnarray}
A_{\parallel} (t) = - E_0 T f \Bigl(\frac{t}{T} \Bigr),
\end{eqnarray}
where $f(t/T)$ is a dimensionless function, and we introduce the dimensionless Keldysh parameter
\begin{eqnarray}
\gamma = \frac{m}{qE_0T}. \label{Keldysh}
\end{eqnarray}
Other useful dimensionless parameters for pulsed electric fields in Ref.~~\cite{Kim:2007pm} are
\begin{eqnarray}
\epsilon_T = \frac{m}{qE_0 \bar{T}}, \quad \delta = \frac{qE_0}{\pi m^2}, \label{KP}
\end{eqnarray}
where
\begin{eqnarray}
E_0 \bar{T} = \frac{1}{2} \int_{- \infty}^{\infty} E(t)dt, \quad E_0 = \mathrm{max}\,(|E(t)|).
\end{eqnarray}
The parameter $\epsilon_T$ in Eq.~(\ref{KP}) is the same as the Keldysh parameter in Eq.~(\ref{Keldysh}) for the Sauter-type electric field $E_0\, \mathrm{sech}^2 (t/T)$ but differs by a numerical factor of $2/\sqrt{\pi}$ for the Gaussian type electric field $E_0\, e^{-(t/T)^2}$. An electric field with two time scales $E_0\, \cos (\omega t)\, e^{- (t/T)^2}$ leads to $\bar{T} = (\sqrt{\pi}/2)\, T\, e^{- (\omega T)^2/4}$.
Then, the symmetrized frequency takes the form
\begin{eqnarray}
\bar{\omega}_{\bf k} = \frac{m}{2} \Biggl[ \sqrt{1 + \kappa^2 + \Bigl( \mu + \frac{f}{\gamma}\Bigr)^2} + \sqrt{1 + \kappa^2 + \Bigl( \mu - \frac{f}{\gamma}\Bigr)^2}  \Biggr],
\end{eqnarray}
where $\kappa$ and $\mu$ are the dimensionless momenta
\begin{eqnarray}
\kappa = \frac{|{\bf k}_{\perp}|}{m}, \quad \mu = \frac{k_{\parallel}}{m}.
\end{eqnarray}
We introduce another useful dimensionless function
\begin{eqnarray}
{\cal F} (z) \equiv \sqrt{1 + \Bigl(\frac{f(z)}{\gamma}\Bigr)^2}, \quad z = \frac{t}{T}. \label{F fun}
\end{eqnarray}
Then the worldline-instanton action is the extremum of the zero-momentum action
\begin{eqnarray}
{\cal S}_{(0)} = i m T \oint {\cal F} (z) dz, \label{wkb 0-act}
\end{eqnarray}
and contributions of power of momenta are given by
\begin{eqnarray}
{\cal F}^{(n)} = i m T \oint \frac{1}{\bigl({\cal F} (z)\bigr)^n} dz.
\end{eqnarray}
Note that ${\cal S}_{(0)}$ and ${\cal F}^{(n)}$ are dimensionless.

Integrating the symmetrized action up through the quadratic momenta
\begin{eqnarray}
2 {\cal S}_{\bf k} = {\cal S}^{(0)} + \frac{1}{2} {\cal F}^{(1)} \kappa^2 + \frac{1}{2} {\cal F}^{(3)} \mu^2 + \cdots \label{mom exp}
\end{eqnarray}
gives the pair production per unit volume
\begin{eqnarray}
\frac{d^3{\cal N}}{dx^3} = \frac{m^3}{(2 \pi)^{3/2}} \frac{e^{- {\cal S}_{(0)}}}{\vert {\cal F}^{(1)} \vert \sqrt{\vert {\cal F}^{(3)} \vert } }.
\end{eqnarray}
In obtaining Eq.~(\ref{mom exp}), the square is expanded through second order in $\mu$, otherwise the coefficient would be proportional to ${\cal F}^{(1)}$.
To calculate ${\cal S}_{(0)}$ and ${\cal F}^{(n)}$, we change the variable motivated by the Euclidean time \cite{Dunne:2006st}
\begin{eqnarray}
y = i \frac{f(z)}{\gamma},
\end{eqnarray}
and obtain
\begin{eqnarray}
{\cal S}_{(0)} = m \gamma T \oint \frac{\sqrt{1- y^2}}{f' (y)} dy = \frac{\pi m^2}{q E_0} g(r^2) \label{world ins}
\end{eqnarray}
where $g$ is the action for the back and forth motion divided by $\pi$,
\begin{eqnarray}
g(r^2) = \frac{2}{\pi} \int_{-1}^{1} \frac{\sqrt{1- y^2}}{f' (y)} dy. \label{per int}
\end{eqnarray}
It should be noted that Eq.~{(\ref{world ins}) is the worldline instanton action in Refs.~\cite{Dunne:2005sx,Dunne:2006st}. Further note that ${\cal F}^{(n)}$ for $n \geq 2$ can be expressed as derivatives of ${\cal F}^{(1)}$ or ${\cal S}_{(0)} = {\cal F}^{(-1)}$ with respect to $ \gamma^2$ through the recursive relation
\begin{eqnarray}
\Bigl( 1+ \frac{f^2}{\gamma^2} \Bigr)^{- \frac{n}{2} -1} = - \frac{2}{n} \gamma^{n+2} \frac{d}{d(\gamma^2)} \Bigl[\gamma^{-n} \Bigl( 1+ \frac{f^2}{\gamma^2} \Bigr)^{- \frac{n}{2}} \Bigr].
\end{eqnarray}

\subsection{Space-dependent electric fields}\label{sec2-a}

Next, consider a space-dependent but static electric field given by a Coulomb potential
\begin{eqnarray}
A_0 (x_{\parallel} ) = - E_0 L f \Bigl(\frac{x_{\parallel}}{L} \Bigr),
\end{eqnarray}
which gives the four-momentum mode equation
\begin{eqnarray}
{\phi}''_{\omega {\bf k}_{\perp}} (x_{\parallel}) + k^2 (x_{\parallel}) \phi_{\omega {\bf k}_{\perp}} (x_{\parallel}) = 0, \label{sc x-eq}
\end{eqnarray}
where primes denote derivatives with respect to $x_{\parallel}$  and the longitudinal momentum is given by
\begin{eqnarray}
k^2 (x_{\parallel}) = \Bigl(\omega - qE_0 L f \bigl(\frac{x_{\parallel}}{L} \bigr) \Bigr)^2 - m^2 - {\bf k}_{\perp}^2.
\end{eqnarray}
There is a tunneling barrier, under which $k^2 (x_{\parallel}) \leq 0$. So, in contrast to Eq.~(\ref{sc eq}) for time-dependent electric fields, Eq. (\ref{sc x-eq}) becomes a tunneling problem, and the relativistic instanton actions were used to calculate the pair production densities \cite{Kim:2000un,Kim:2003qp,Kim:2007pm}. The phase-integral
$\phi_{\omega {\bf k}_{\perp}} (x_{\parallel}) = e^{i {\cal S}_{\omega {\bf k}_{\perp}}}$ outside the barrier gives the WKB instanton action for the tunneling probability under the barrier in the complex plane $z$,
\begin{eqnarray}
{\cal S}_{\omega {\bf k}_{\perp}} = L \oint \bar{k} (z) dz,
\end{eqnarray}
where
\begin{eqnarray}
\bar{k}(z) = \frac{1}{2} \Biggl[\sqrt{m^2 + {\bf k}_{\perp}^2 - \Bigl(\omega + \frac{m}{\gamma_L} f (z) \Bigr)^2  } + \sqrt{m^2 + {\bf k}_{\perp}^2 - \Bigl(\omega - \frac{m}{\gamma_L} f (z) \Bigr)^2} \Biggl].
\end{eqnarray}
Here, a dimensionless parameter analogous to the Keldysh parameter in Eq.~(\ref{Keldysh}) is introduced,
\begin{eqnarray}
\gamma_l = \frac{m}{qE_0L}.
\end{eqnarray}

Using the dimensionless momenta and energy
\begin{eqnarray}
\kappa = \frac{|{\bf k}_{\perp}|}{m}, \quad \mu = \frac{\omega}{m},
\end{eqnarray}
we may write
\begin{eqnarray}
\bar{k}(z) = \frac{m}{2} \Biggl[ \sqrt{1 + \kappa^2 - \Bigl( \mu + \frac{f}{\gamma_l}\Bigr)^2} + \sqrt{1 + \kappa^2 - \Bigl( \mu - \frac{f}{\gamma_l}\Bigr)^2}  \Biggr].
\end{eqnarray}
As in the case of time-dependent electric fields, we introduce another useful function
\begin{eqnarray}
{\cal G} (z) \equiv \sqrt{1 - \Bigl(\frac{f(z)}{\gamma}\Bigr)^2}, \quad z = \frac{x_{\parallel}}{L},
\end{eqnarray}
and then the worldline instanton is the tunneling action with zero momentum,
\begin{eqnarray}
{\cal S}_{(0)} =  m L \oint {\cal G} (z) dz,
\end{eqnarray}
and contributions of powers of momenta are given by
\begin{eqnarray}
{\cal G}^{(n)} =  m L \oint \frac{dz}{\bigl({\cal G} (z)\bigr)^n}.
\end{eqnarray}
The integration of momenta and energy up to through quadratic order gives the pair production per unit time and per unit transverse area,
\begin{eqnarray}
\frac{d^3{\cal N}}{dx^3} = \frac{m^3}{(2 \pi)^{3/2}} \frac{e^{- {\cal S}_{(0)}}}{\vert {\cal G}^{(1)} \vert \sqrt{\vert {\cal G}^{(3)} \vert } }.
\end{eqnarray}

\subsection{Sauter-type and sinusoidal electric fields}\label{sec2-b}

As specific models for pair production density, we consider a Sauter-type electric field $E(t) = E_0 \, {\rm sech}^2 (t/T)$ and a sinusoidal electric field $E (t) = E_0 \cos (t/T)$, for which the dimensionless gauge functions are $f(z) = \tanh (z)$ and $f(z) = \sin(z)$, respectively. The complex analysis in Ref.~\cite{Kim:2007pm} will be used below.

First, for $f(z) = \tanh (z)$, we adopt the conformal mapping $\zeta = \tanh(z)$ and introduce a branch cut connecting $i \gamma$ and $-i \gamma$, which makes the square root an analytic function. Then, the WKB action (\ref{wkb 0-act}) along a clockwise contour enclosing the simple pole $\zeta = 0$ receives a series of residues
\begin{eqnarray}
{\cal S}_{(0)} &=& i m T \oint  \frac{d \zeta}{1- \zeta^2} \sqrt{1 + \frac{\zeta^2}{\gamma^2}} \nonumber\\
&=& 2 \pi \frac{mT}{\gamma} \Biggl[ \sum_{n = 0}^{\infty} {\frac{1}{2} \choose n} \gamma^{2n} - 1 \Biggr] \nonumber\\
&=& \frac{\pi m^2}{qE_0} \Biggl( \frac{2}{ 1+ \sqrt{1+ \gamma^2}} \Biggr),
\label{wkb saut}
\end{eqnarray}
where we expanded the square root in a power series in $\gamma/\zeta$ and $1/(1- \zeta^2)$ in a power series in $\zeta$.
The next leading terms for the WKB action for $p =1, 3$ give the prefactors,
\begin{eqnarray}
{\cal F}^{(p)} &=& i m T \oint  \frac{d \zeta}{1 - \zeta^2} \Bigl(1 + \frac{\zeta^2}{\gamma^2}\Bigr)^{- \frac{p}{2}} \nonumber\\
&=& 2 \pi mT \gamma^p  \sum_{n = 0}^{\infty} {- \frac{p}{2} \choose n} \gamma^{2n}  \nonumber\\
&=& 2 \pi mT \gamma^p \Bigl(1 + \gamma^2\Bigr)^{- \frac{p}{2}}. \label{saut pref}
\end{eqnarray}
Note that ${\cal F}^{(p)} = 0$ for even integers $p$. Therefore, the pair production density is
\begin{eqnarray}
\frac{d^3 {\cal N}}{dx^3} = \frac{(qE_0)^{5/2} T}{(2 \pi)^3 m} (1+ \gamma^2)^{5/4} \exp \Bigl[- \frac{\pi m^2}{qE_0} \Bigl( \frac{2}{ 1+ \sqrt{1+ \gamma^2}} \Bigr)\Bigr], \label{saut E}
\end{eqnarray}
which is the same as the result from the phase-integral \cite{Kim:2007pm} and the worldline instanton action combined with the prefactor \cite{Dunne:2005sx,Dunne:2006st}.

Second, for $f(z) = \sin (z)$, we use the conformal mapping $\zeta = \sin(z)$ and write the leading WKB action as
\begin{eqnarray}
{\cal S}_{(0)} = i m T \oint  d \zeta \sqrt{ \frac{1 + \frac{\zeta^2}{\gamma^2}}{1 - \zeta^2}}.
\label{wkb sin}
\end{eqnarray}
Expanding the square root $(1 + \zeta^2/\gamma^2)^{1/2}$ in a power series of $\gamma/\zeta$ and $1/(1- \zeta^2)^{1/2}$ in a power series of $\zeta$ and taking a clockwise contour enclosing $\zeta = 0$, we obtain
\begin{eqnarray}
{\cal S}_{(0)} &=& \pi m T \gamma \sum_{n= 0}^{\infty} \frac{(-1)^n}{n+1} \Bigg[{- \frac{1}{2} \choose n} \gamma^n \Biggr]^2 \nonumber\\
&=& \frac{\pi m^2}{qE_0} F \Bigl(\frac{1}{2}, \frac{1}{2}; 2; - \gamma^2 \Bigr), \label{saut act}
\end{eqnarray}
where $F$ is the hypergeometric function. Using the recursion relation 15.2.25 for hypergeometric functions in Ref.~\cite{Abramowitz:1964}, the series representation of the complete elliptic integrals of the first kind $K$ and the second kind $E$ and the relation 8.127 in Ref.~\cite{Gradshteyn:1994}, we write the leading WKB action in another form
\begin{eqnarray}
{\cal S}_{(0)} &=& \frac{\pi m^2}{qE_0} \times \frac{4}{\pi \gamma^2} \bigl[ (1+ \gamma^2) K(- i \gamma) - E(- i \gamma) \bigr] \nonumber\\
 &=& \frac{4 m^2}{qE_0} \frac{\sqrt{1+\gamma^2}}{\gamma^2} \Bigl[K \Bigl(\frac{\gamma}{\sqrt{1+ \gamma^2}} \Bigr) - E \Bigl(\frac{\gamma}{\sqrt{1+ \gamma^2}} \Bigr) \Bigr]. \label{ell int}
\end{eqnarray} 
A direct integration of Eq.~(\ref{per int}) confirms the result (\ref{ell int}). Note that the leading WKB action is the same as the worldline instanton action in Refs.~\cite{Dunne:2005sx,Dunne:2006st} Similarly, using the relations \cite{series} and the recursion relation 9.137 for hypergeometric functions in Ref.~\cite{Gradshteyn:1994}, we obtain the prefactors
\begin{eqnarray}
{\cal F}^{(1)} &=& \frac{2 \pi m^2}{qE_0} \times \frac{2}{\pi} K(-i \gamma) \nonumber\\
&=& \frac{2 \pi m^2}{qE_0} \times \frac{2}{\pi \sqrt{1+ \gamma^2}} K \Bigl(\frac{\gamma}{\sqrt{1+ \gamma^2}} \Bigr), \nonumber\\
{\cal F}^{(3)} &=& \frac{2 \pi m^2}{qE_0} \times  \gamma^2 F \Bigl(\frac{3}{2}, \frac{3}{2}; 2; - \gamma^2 \Bigr) \nonumber\\
&=& \frac{2 \pi m^2}{qE_0} \times \frac{4}{\pi \sqrt{1+ \gamma^2}} \Bigl[K \Bigl(\frac{\gamma}{\sqrt{1+ \gamma^2}} \Bigr) - E \Bigl(\frac{\gamma}{\sqrt{1+ \gamma^2}} \Bigr) \Bigr].
\end{eqnarray}
Then, the pair production density is
\begin{eqnarray}
\frac{d^3 {\cal N}}{dx^3} &=& \frac{(qE_0)^{5/2} T}{16 \pi^2 m} \frac{ \exp \Bigl[ - \frac{\pi m^2}{qE_0} F \Bigl(\frac{1}{2}, \frac{1}{2}; 2; - \gamma^2 \Bigr)\Bigr]}{ K(- i \gamma) \sqrt{ F \Bigl(\frac{3}{2}, \frac{3}{2}; 2; - \gamma^2 \Bigr)}} \nonumber\\
&=& \frac{(qE_0)^{3/2} (1+ \gamma^2)^{3/4}}{32 \pi^{3/2}} \frac{\exp \Bigl[-\frac{4 m^2}{qE_0} \frac{\sqrt{1+\gamma^2}}{\gamma^2} \Bigl\{K \Bigl(\frac{\gamma}{\sqrt{1+ \gamma^2}} \Bigr) - E \Bigl(\frac{\gamma}{\sqrt{1+ \gamma^2}} \Bigr) \Bigr\} \Bigr]}{ K \Bigl(\frac{\gamma}{\sqrt{1+ \gamma^2}} \Bigr) \sqrt{ K \Bigl(\frac{\gamma}{\sqrt{1+ \gamma^2}} \Bigr) - E \Bigl(\frac{\gamma}{\sqrt{1+ \gamma^2}} \Bigr) }}.
\end{eqnarray}

Next, we consider the space-dependent but static electric field $E(x_{\parallel}) = E_0 \, {\rm sech}^2 (x_{\parallel}/L)$ and a sinusoidal electric field $E (x_{\parallel}) = E_0 \cos (x_{\parallel}/L)$. The dimensionless gauge functions are the same as the homogeneous time-dependent fields $f(z) = \tanh (z)$ and $f(z) = \sin(z)$. In the tunneling region for $E(x_{\parallel}) = E_0 \, {\rm sech}^2 (x_{\parallel}/L)$, we analytically continue and write the leading WKB action as
\begin{eqnarray}
{\cal S}_{(0)} = i m L \oint  \frac{d \zeta}{1- \zeta^2} \sqrt{\frac{\zeta^2}{\gamma_l^2}-1},
\end{eqnarray}
and
\begin{eqnarray}
{\cal G}^{(p)} = e^{- i \pi p/2} m L \oint  \frac{d \zeta}{1 - \zeta^2} \Bigl(\frac{\zeta^2}{\gamma_l^2} - 1\Bigr)^{- \frac{p}{2}}.
\end{eqnarray}
Repeating the procedure for the time-dependent case, we obtain the pair production density
\begin{eqnarray}
\frac{d^3 {\cal N}}{dx^3} = \frac{(qE_0)^{5/2} L}{(2 \pi)^3 m} (1- \gamma_l^2)^{5/4} \exp \Bigl[- \frac{\pi m^2}{qE_0} \Bigl( \frac{2}{ 1+ \sqrt{1- \gamma_l^2}} \Bigr)\Bigr].
\end{eqnarray}

Similarly, the complex analysis for $E (x_{\parallel}) = E_0 \cos (x_{\parallel}/L)$ leads to the pair production density per unit transverse area and per unit longitudinal period
\begin{eqnarray}
\frac{d^3 {\cal N}}{dx^3} = \frac{(qE_0)^{5/2} L}{(2 \pi)^3 m} \frac{ \exp \Bigl[ - \frac{\pi m^2}{qE_0} F \Bigl(\frac{1}{2}, \frac{1}{2}; 2;  \gamma_l^2 \Bigr)\Bigr]}{\Bigl| \frac{2 \gamma_l}{\pi} K(\gamma_l^2) \Bigr| \sqrt{\Bigl| F \Bigl(\frac{3}{2}, \frac{3}{2}; 2; \gamma_l^2 \Bigr) \Bigr|}} .
\end{eqnarray}
The energetic condition requires $\gamma_l \leq 1$ for pair production in space-dependent electric fields. There is an analytical continuation of $\gamma \leftrightarrow i \gamma_l$ between the time-dependent and space-dependent electric fields studied in this section. The leading WKB action ${\cal S}_{(0)}$ and the factors ${\cal G}^{(1)}$ and ${\cal G}^{(3)}$ are larger for a space-dependent electric field than those for a time-dependent electric field of the same profile function and therefore the pair production density is exponentially and polynomially suppressed.

\section{Fermion production}\label{sec3}

In a spatially homogeneous but time-dependent electric field, the Dirac equation for spin-1/2 fermions gives the Fourier-momentum mode equation for each spin
\begin{eqnarray}
\ddot{\phi}_{{\bf k} \sigma} (t) + \omega_{{\bf k} \sigma}^2 (t) \phi_{{\bf k} \sigma} (t) = 0, \label{sp eq}
\end{eqnarray}
where the spin-dependent frequency for each spin $\sigma = \pm 1/2$ is
\begin{eqnarray}
\omega_{{\bf k} \sigma}^2 (t) =  m^2 + {\bf k}_{\perp}^2 + \bigl(k_{\parallel} - qA_{\parallel} (t) \bigr)^2 + 2 i \sigma qE (t).
\end{eqnarray}
The last imaginary term comes from the diagonal components of the spin-tensor of the second-order Dirac equation.
As for the scalar case, the square root of the product of the sum of $({\bf k}, \sigma)$ and another sum of $(-{\bf k}, - \sigma)$ gives the symmetrized frequency in the form
\begin{eqnarray}
\bar{\omega}_{{\bf k} \sigma} = \frac{m}{2} \Biggl[ \sqrt{1 + \kappa^2 + \Bigl( \mu + \frac{f}{\gamma}\Bigr)^2 + 2  \pi i
\sigma \delta f'} + \sqrt{1 + \kappa^2 + \Bigl( \mu - \frac{f}{\gamma}\Bigr)^2 - 2 \pi i \sigma \delta f'}  \Biggr], \label{fer freq}
\end{eqnarray}
where $\delta = q E_0/ \pi m^2$ and primes denote the derivatives with respect to $z = t/T$. Note that $\bar{\omega}_{-{\bf k} -\sigma} =\bar{\omega}_{{\bf k} \sigma}$, as expected.

We follow the scalar case by simply modifying the functional ${\cal F}$ to become
\begin{eqnarray}
{\cal F}_{\sigma} = \sqrt{1 + \Bigl(\frac{f}{\gamma}\Bigr)^2 + 2 \pi i \sigma \delta f'},
\end{eqnarray}
and expand Eq.~(\ref{fer freq}) up to quadratic order of momenta
\begin{eqnarray}
\bar{\omega}_{{\bf k} \sigma} = \frac{m}{2} \bigl( {\cal F}_{\sigma} + {\cal F}_{\sigma}^* \bigr) + \frac{m}{4} \Bigl[ \Bigl(
\frac{1}{{\cal F}_{\sigma}} + \frac{1}{{\cal F}_{\sigma}^*}\Bigr) \kappa^2 + \Bigl(
\frac{1+ 2 \pi i  \sigma \delta f'}{({\cal F}_{\sigma})^3} + \frac{1-2 \pi i \sigma \delta f'}{({\cal F}_{\sigma}^*)^3}\Bigr) \mu^2  +
2 \mu \frac{f}{\gamma} \Bigl(
\frac{1}{{\cal F}_{\sigma}} - \frac{1}{{\cal F}_{\sigma}^*}\Bigr) \Bigr]. \label{fer 2nd}
\end{eqnarray}
In the phase-integral method, the WKB action $i T \oint \bar{\omega}_{{\bf k} \sigma} (z) dz$ gets residues from poles in the complex plane of time and become real and positive except for the last term in the square bracket of Eq.~(\ref{fer 2nd}) which gives a phase factor and thus does not contribute the pair production.
Then, the leading WKB action is the zero-momentum action
\begin{eqnarray}
{\cal S}_{(0)\sigma} = i \frac{mT}{2} \oint \Bigl( {\cal F}_{\sigma} (z) + {\cal F}_{\sigma}^* (z) \Bigr)dz,
\end{eqnarray}
and contributions of powers of momenta are given by
\begin{eqnarray}
\bar{\cal F}^{(1)}_{\sigma} = i \frac{mT}{2} \oint \Bigl(
\frac{1}{{\cal F}_{\sigma}} + \frac{1}{{\cal F}_{\sigma}^*}\Bigr) dz,
\end{eqnarray}
and
\begin{eqnarray}
\tilde{\cal F}^{(3)}_{\sigma} = i \frac{mT}{2} \oint \Bigl(
\frac{1+ 2 \pi i \sigma \delta f'}{({\cal F}_{\sigma})^3} + \frac{1-2 \pi i \sigma \delta f'}{({\cal F}_{\sigma}^*)^3}\Bigr) dz.
\end{eqnarray}
Finally, we obtain the fermion production density
\begin{eqnarray}
\frac{d{\cal N}_{\rm sp}}{dx^3} = \frac{m^3}{(2 \pi)^{3/2}} \sum_{\sigma} \frac{e^{- {\cal S}_{(0) \sigma}}}{\vert \bar{\cal F}^{(1)}_{\sigma} \vert \sqrt{\vert \tilde{\cal F}^{(3)}_{\sigma} \vert } }.
\end{eqnarray}
The WKB action and the prefactors are already symmetrized with respect to $\sigma$ and $- \sigma$, so the sum of spin components just gives a factor of 2.

As an illustration, we consider the Sauter-type electric field $E(t) = E_0 \, {\rm sech}^2 (t/T)$. We introduce a dimensionless parameter for spin-1/2 fermions
\begin{eqnarray}
\gamma_{\sigma} = \gamma \sqrt{\frac{1 + 2 \pi i \sigma \delta}{1 - 2 \pi i \sigma \delta \gamma^2}},
\end{eqnarray}
and write
\begin{eqnarray}
{\cal F}_{\sigma} (z) = \sqrt{1 + 2 \pi i \sigma \delta} \sqrt{1+ \Bigl(\frac{f(z)}{\gamma_{\sigma}} \Bigr)^2}.
\end{eqnarray}
Then, following Sec.~\ref{sec2-b}, the leading WKB action is given by
\begin{eqnarray}
{\cal S}_{(0)}^{\rm sp} = \frac{\pi m^2}{qE_0} \Bigl(\frac{1+ i \pi \delta}{\sqrt{1+ \gamma^2} + \sqrt{1 - i \pi \delta \gamma^2}} + \frac{1 - i \pi \delta}{\sqrt{1+ \gamma^2} + \sqrt{1 + i \pi \delta \gamma^2}} \Bigr), \label{saut fer}
\end{eqnarray}
and the prefactors are independent of the spin components,
\begin{eqnarray}
\bar{\cal F}^{(1)} &=& \frac{2 \pi m^2}{qE_0} \Bigl(1+ \gamma^2 \Bigr)^{- \frac{1}{2}}, \nonumber\\
\tilde{\cal F}^{(3)} &=& \frac{2 \pi m^2}{qE_0} \Bigl(1+ \gamma^2 \Bigr)^{- \frac{3}{2}}. \label{fer pref}
\end{eqnarray}
It should be noted that the integrals over a complete period in the complex plane of time give the same result as Eqs.~(\ref{saut fer}) and (\ref{fer pref}). Finally, summing the spin components, the pair production density is given by
\begin{eqnarray}
\frac{d^3 {\cal N}^{\rm sp}}{dx^3} = \frac{2 (qE_0)^{5/2} T}{(2 \pi)^3 m} (1+ \gamma^2)^{5/4} \exp \Bigl[- \frac{\pi m^2}{qE_0} \Bigl(\frac{1+ i \pi \delta}{\sqrt{1+ \gamma^2} + \sqrt{1 - i \pi \delta \gamma^2}} + \frac{1 - i \pi \delta}{\sqrt{1+ \gamma^2} + \sqrt{1 + i \pi \delta \gamma^2}} \Bigr)\Bigr].
\end{eqnarray}

\section{Parallel constant magnetic field}\label{sec4}

In a homogeneous, time-dependent electric field and a parallel constant magnetic field, the charged scalar field has the time-dependent frequency
\begin{eqnarray}
\omega_{k_{\parallel} n}^2 (t) =  m^2 + qB_0 (2n+1) + \bigl(k_{\parallel} - qA_{\parallel} (t) \bigr)^2.
\end{eqnarray}
Then the symmetrized frequency takes the form
\begin{eqnarray}
\bar{\omega}_{k_{\parallel}n} = \frac{m}{2} \Bigl[ \sqrt{1 + \pi \delta_B (2n+1) + \Bigl( \mu + \frac{f}{\gamma}\Bigr)^2} + \sqrt{1 + \pi \delta_B (2n+1) + \Bigl( \mu - \frac{f}{\gamma}\Bigr)^2}  \Bigr],
\end{eqnarray}
where $\delta_B = qB_0/\pi m^2$.
The leading WKB action is given by the integral of the dimensionless function (\ref{F fun}),
\begin{eqnarray}
{\cal S}_{(0)} = i m T \oint {\cal F} (z) dz,
\end{eqnarray}
and contributions of powers of momenta or Landau levels have the same form
\begin{eqnarray}
{\cal F}^{(n)} = i m T \oint \frac{dz}{({\cal F} (z))^n}.
\end{eqnarray}
Note that ${\cal S}_{(0)}$ and ${\cal F}^{(n)}$ are dimensionless.

In the weak magnetic field regime $(\delta_B < 1)$, integrating the symmetrized action up through the quadratic momentum and summing the Landau levels,
\begin{eqnarray}
2 {\cal S}_{\bf k} = {\cal S}^{(0)} + \frac{1}{2} {\cal F}^{(1)} \pi \delta_B (2n+1) + \frac{1}{2} {\cal F}^{(3)} \mu^2 + \cdots, \label{B-mom exp}
\end{eqnarray}
gives the pair production per unit volume
\begin{eqnarray}
\frac{d^3{\cal N}}{dx^3} = \frac{m qB_0}{2 (2 \pi)^{3/2}} \frac{e^{- {\cal S}_{(0)}}}{\vert \sinh \Bigl(\frac{\pi \delta_B}{2}  {\cal F}^{(1)} \Bigr) \vert \sqrt{\vert {\cal F}^{(3)} \vert } }.
\end{eqnarray}

In the Sauter-type electric field $E(t) = E_0 \, {\rm sech}^2 (t/T)$, the leading WKB action ${\cal S}_{(0)}$ is the same as Eq.~(\ref{wkb saut}) and the prefactors ${\cal F}^{(1)}$ and ${\cal F}^{(3)}$ are the same as Eq.~(\ref{saut pref}). Therefore, the pair production density is
\begin{eqnarray}
\frac{d^3 {\cal N}}{dx^3} = \frac{(qB_0) (qE_0)^{3/2} T}{2 (2 \pi)^2 m}  \frac{(1+ \gamma^2)^{3/4}}{\sinh \Bigl(\frac{B_0}{E_0} \frac{\pi}{\sqrt{1+ \gamma^2}} \Bigr)}
 \exp \Bigl[- \frac{\pi m^2}{qE_0} \Bigl( \frac{2}{ 1+ \sqrt{1+ \gamma^2}} \Bigr)\Bigr]. \label{saut B}
\end{eqnarray}
In the limit of zero magnetic field, the pair production density (\ref{saut B}) reduces to that in Eq.~(\ref{saut E}).

\section{Conclusion}\label{sec5}

We have extended the phase-integral method for Schwinger pair production in Ref.~\cite{Kim:2007pm} to any electric field of fixed direction
with value depending on only one-time or space coordinate and showed that the  momenta integration up through quadratic order recovers the pair production density from the worldline instanton including the prefactor \cite{Dunne:2005sx,Dunne:2006st}. The physical reasoning behind this equivalence is that the phase-integral is the WKB action from relativistic field equations and includes quantum fluctuations in momenta whereas the worldline instanton is the action of the charge along the electric field in the Euclidean spacetime and then the prefactor takes into account quantum fluctuations. As illustrations, we have found the pair production densities both in Sauter-type and sinusoidal time-dependent or space-dependent electric fields.

We have further extended the phase-integral method to fermion production in any such time-dependent electric field. The phase-integral method for fermions has one distinct feature: an imaginary time-dependent electric field with spin components from the spin-tensor of the second order Dirac equation. We applied the phase-integral method to a time-dependent electric field with a parallel constant magnetic field and obtained the pair-production density. The pair production density for fermions has been found for the Sauter-type time-dependent electric field and for charged spinless bosons in both the Sauter-type time-dependent electric field and a parallel constant magnetic field.

\acknowledgments
S.P.K. would like to thank the warm hospitality at the University of Alberta and Rong-Gen Cai for the warm hospitality at the Institute of Theoretical Physics (ITP), Chinese Academy of Sciences (CAS).
The work of S.P.K. was supported in part by the Institute for Basic Science (IBS) under IBS-R012-D1 and in part by the Open Project Program of State Key Laboratory of Theoretical Physics, Institute of Theoretical Physics, Chinese Academy of Sciences, China (No.~Y5KF161CJ1) and the Fundamental Research Funds for the Central Universities. The work of D.N.P. was supported by the Natural Sciences and Engineering Research Council of Canada.

\end{document}